\documentclass[revtex4]{emulateapj}
\usepackage{natbib}
\usepackage{graphicx}
\usepackage{multirow}
\usepackage{arydshln}
\usepackage{txfonts}
\usepackage{appendix}
\usepackage{bigdelim}
\usepackage{soul,ulem}
\newcommand{\kms}{\mbox{km~s$^{-1}$}}

\newcommand{\rs}{\mbox{$R_{\star}$}}

\newcommand{\vlsr}{\mbox{$V_{\mathrm{LSR}}$}}

\newcommand{\arcsecp}{\mbox{\rlap{.}$''$}} 
\newcommand{\secp}{\mbox{\rlap{.}$^{\mathrm{s}}$}} 
\newcommand{\jyperb}{\mbox{Jy\,beam$^{-1}$}}

\begin{document}
%% TITLE: ----------------------------------------------------------

\title{Si-bearing molecules toward IRC+10216: ALMA unveils the molecular envelope of CWLeo}

%% AUTHORS AND AFFIL.: ----------------------------------------------------------

\author{L.~Velilla Prieto\altaffilmark{1,2}, 
J.~Cernicharo\altaffilmark{1}, 
G.~Quintana--Lacaci\altaffilmark{1},
M.~Ag\'{u}ndez\altaffilmark{1},
A.~Castro--Carrizo\altaffilmark{3},
J.~P.~Fonfr\'{i}a\altaffilmark{4},
N.~Marcelino\altaffilmark{5},
J.~Z\'{u}\~niga\altaffilmark{6},
A.~Requena\altaffilmark{6},
A.~Bastida\altaffilmark{6},
F.~Lique\altaffilmark{7} and
M.~Gu\'{e}lin\altaffilmark{3,8} 
}
\altaffiltext{1}{\footnotesize{Group of Molecular Astrophysics. ICMM, CSIC. C/ Sor Juana In\'es de la Cruz 3, 28049 Cantoblanco, Madrid, Spain}}
\altaffiltext{2}{\footnotesize{Centro de Astrobiolog\'ia, INTA--CSIC. E-28691 Villanueva de la Ca\~nada, Madrid, Spain}}
\altaffiltext{3}{\footnotesize{Institut de Radioastronomie Millim\'{e}trique. 300 rue de la la Piscine, F-38406, Saint Martin d'H\`{e}res, France}}
\altaffiltext{4}{\footnotesize{Departamento de Estrellas y Medio Interestelar, Instituto de Astronom\'{i}a, Universidad Nacional Aut\'{o}noma de M\'{e}xico (UNAM). Ciudad Universitaria, 04510, Mexico City, M\'{exico}}}
\altaffiltext{5}{\footnotesize{Istituto di Radiastronomia, INAF--CNR, via Gobetti 101, 40129 Bologna, Italy}}
\altaffiltext{6}{\footnotesize{Departamento de Qu\'imica F\'isica, Facultad de Qu\'imica, Universidad de Murcia. Campus Espinardo, 30100 Murcia, Spain}}
\altaffiltext{7}{\footnotesize{LOMC--UMR 6294, CNRS--Universit\'{e} du Havre, 25 rue Philippe Lebon, BP. 1123, 76063 Le Havre cedex, France}}
\altaffiltext{8}{\footnotesize{LERMA, Observatoire de Paris, PSL Research University, CNRS, UMR 8112, F-75014, Paris, France}}

\journalinfo{Published in ApJL 2015 May 27 // 2015ApJ...805L..13V} 
\submitted{Submitted 2015 April 8 --- Published in ApJL 2015 May 27}

%% ABSTRACT: ----------------------------------------------------------

\begin{abstract}
We report the detection of SiS rotational lines in high-vibrational states as well as SiO and SiC$_2$ lines in their ground vibrational state toward IRC+10216 during the Atacama Large Millimeter Array Cycle\,0.
The spatial distribution of these molecules shows compact emission for SiS and a more extended emission for SiO and SiC$_2$, and also proves the existence of an increase in the SiC$_2$ emission at the outer shells of the circumstellar envelope.
We analyze the excitation conditions of the vibrationally excited SiS using the population diagram technique, and we use a large velocity gradient model to compare with the observations.
We found moderate discrepancies between the observations and the models that could be explained if SiS lines detected are optically thick.
Additionally, the line profiles of the detected rotational lines in the high energy vibrational states show a decreasing linewidth with increasing energy levels.
This may be evidence that these lines could be excited only in the inner shells, i.e., the densest and hottest, of the circumstellar envelope of IRC+10216.
\end{abstract}

%% KEYWORDS: ----------------------------------------------------------


%% PAPER: ----------------------------------------------------------
\section{Introduction}\label{sec:intro}
Silicon is mostly locked in SiS, SiO and SiC$_{2}$ in the circumstellar envelope (CSE) of the carbon-rich star IRC+10216, as evidenced observationally and predicted by models \citep{olo82,luc95,agu12}.
These molecules are efficiently formed in the gas phase, close to the stellar photosphere as a consequence of chemical processes enabled under thermodynamical equilibrium \citep{tsu73}.
In the dust formation region ($\sim$5--20\,\rs), the Si-bearing molecules are likely to condense onto the dust grains due to their highly refractory nature.
The silicon contained in the dust grains can form molecules through grain-surface reactions.
Also, the interaction of shocks produced by the pulsation of the star with the dust grains can extract certain amounts of silicon from the grains and incorporate that silicon into the gas-phase to react and form other species \citep[see e.g.][]{cas01}.
Beyond this region, the abundances of Si-bearing molecules are expected to decrease up to the outermost shells of the envelope, where the interstellar ultraviolet (UV) radiation field dissociates all the remaining molecules.

Previous interferometer observations showed the spatial distribution of these molecules in IRC+10216.
The SiS $J$=5 -- 4, $J$=6 -- 5, $J$=8 -- 7, $J$=9 -- 8 and $J$=12 -- 11 brightness distributions display a quasi-circular shape with a diameter of $\sim$20\arcsec\ elongated along the nebular axis \citep[P.A.$\sim$20$^{\circ}$,][]{bie93,luc95}.
Recent observations with the Combined Array for Research in Millimeter-wave Astronomy (CARMA) of the SiS $J$=14 -- 13 v=0 and v=1 lines have been reported by \cite{fon14}, where the v=0 line shows a circular and compact brightness distribution of $\sim$2\arcsec\ and displays maser emission. 
The v=1 brightness distribution shows a compact source centered at the star position.

SiO $J$=5 -- 4 v=0 brightness distribution maps carried out with the Submillimeter Array (SMA) were reported in \cite{sch06}.
They show circular symmetry with a diameter of $\sim$6\arcsec\ at the systemic velocity of the source, which is -26.5\kms\ \citep[e.g.][]{cer00}.
The $J$=6 -- 5 v=0 brightness distribution reported in \cite{fon14} displays a quasi-circular symmetry centered at the position of the star, with a diameter of $\sim$3\arcsec\ elongated along the nebular direction (NE--SW).  

SiC$_2$ observations carried out with the Plateau de Bure Interferometer (PdBI) and CARMA, show a brightness distribution composed of:
($i$) an elongated compact component located at the innermost regions of the CSE  \citep {fon14} and,
($ii$) a hollow shell structure located at $\sim$15\arcsec\ from the star \citep{luc95}.
The formation mechanism for this outer component was suggested in \cite{cer10}, where the reaction between Si and C$_2$H$_2$ yielding SiC$_2$, could be responsible for the SiC$_2$ enhancement in the outer envelope.

In this work we present the Cycle\,0 observations carried out with the Atacama Large Millimeter Array (ALMA) toward IRC+10216.
We detected emission of SiS $J$=15 -- 14 lines of vibrationally excited states, from v=0 up to v=7, and tentatively of v=8, 9 and 10. 
$J$=15 -- 14 lines of different isotopologues are also detected: $^{29}$SiS (v=0--5), $^{30}$SiS (v=0--4), Si$^{33}$S (v=0--3), Si$^{34}$S (v=0--4), $^{29}$Si$^{33}$S (v=0) and $^{29}$Si$^{34}$S (v=0).
We also detected emission of SiO $J$=6 -- 5 (v=0--2), $^{29}$SiO $J$=6 -- 5 (v=0) and several lines of SiC$_2$ in the ground vibrational state.

\begin{figure*}[hbtp!] 
\centering
\includegraphics[width=0.80\hsize]{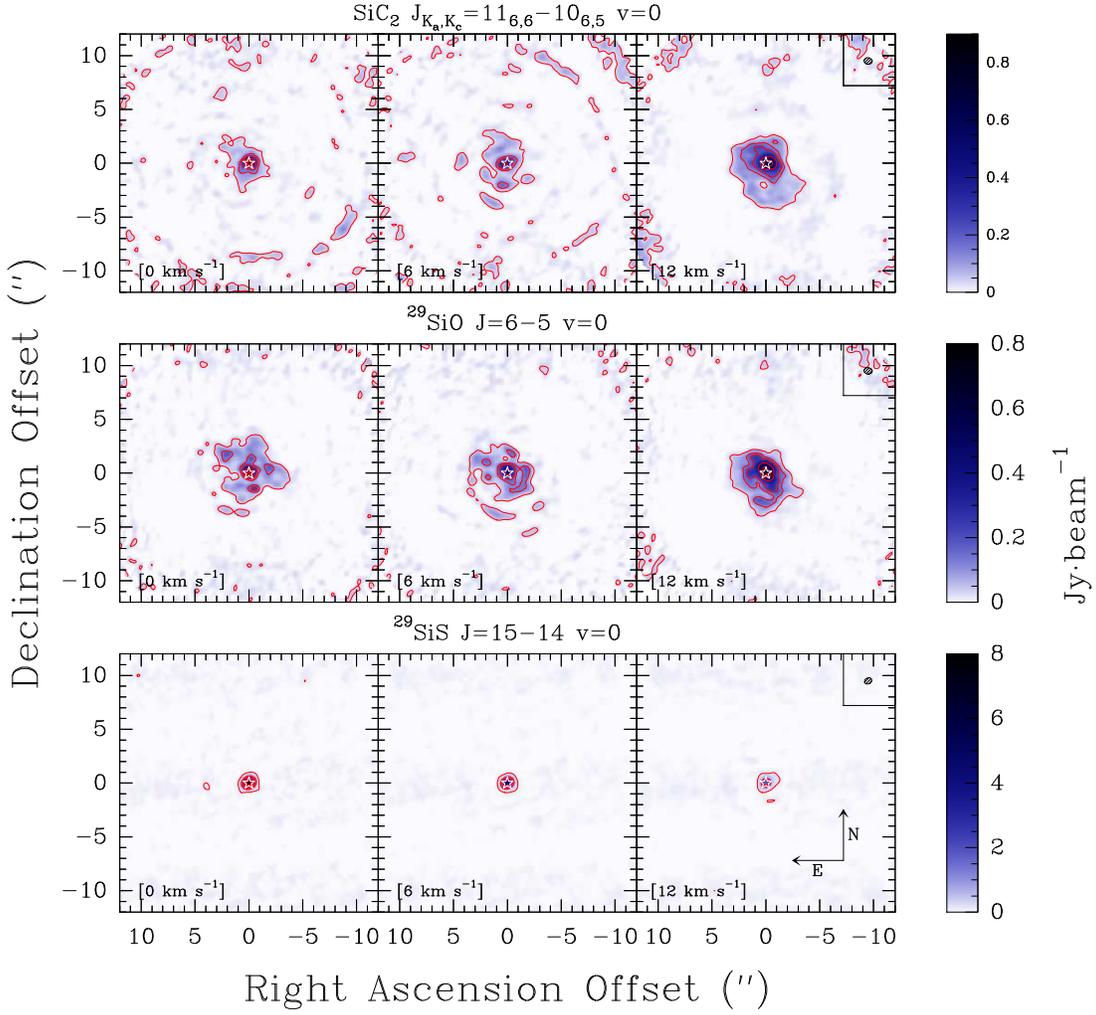}
\caption{From top to bottom: SiC$_2$ $J_{K_{\mathrm{a}},K_{\mathrm{c}}}$=11$_{6,6}$ -- 10$_{6,5}$ v=0, $^{29}$SiO $J$=6 -- 5 v=0 and $^{29}$SiS $J$=15 -- 14 v=0. 
Three velocity offsets with respect to the \vlsr\ of the source are shown: +0, +6 and +12\,\kms.
The position of the star is shown as a white star. 
The first contour corresponds to 5\,$\sigma$ (see Table\,\ref{tab:obs}) and the rest correspond to 15, 25, 50, 75 and 90\%\ of the maximum intensity, except for $^{29}$SiS $J$=15--14 where the first contour is at 25\,$\sigma$ level (a large-scale artificial modulation appears below that level).
The synthetic beam is represented in the top right corner of each of the three maps series.
The intensity scale in \jyperb\ is shown at the right edge of the figure.
The orientation is explicitly shown in the bottom right box.
}
\label{fig:si-bear}
\end{figure*}

\newpage
\section{Observations}\label{sec:obs}
The observations were carried out with ALMA\footnote{\small{This paper makes use of the following ALMA data: ADS/JAO.ALMA\#2011.0.00229.S . ALMA is a partnership of ESO (representing its member states), NSF (USA) and NINS (Japan), together with NRC (Canada), NSC and ASIAA (Taiwan), and KASI (Republic of Korea), in cooperation with the Republic of Chile. The Joint ALMA Observatory is operated by ESO, AUI/NRAO and NAOJ.}}
between 2012 April 8 and 23 during Cycle\,0.
IRC+10216 was observed in the frequency range 255.3 to 274.8\,GHz (band 6) covered by four different setups with a bandwidth of $\sim$5\,GHz, a channel spacing of 0.49\,MHz and an effective resolution of 0.98\,MHz.
Detailed information of each setup is summarized in Table\,\ref{tab:obs}.
The observations were performed using sixteen antennas covering baselines up to 402\,m that allowed us to obtain an angular resolution of $\sim$0.6\arcsec. 
The shortest baselines used were $\sim$20\,m which allow us to recover structures with a size up to $\lesssim$12\arcsec.
Two runs of 72 minutes each were performed, of which 26 minutes correspond to correlations on source.
Further details about calibration and imaging restoration can be found in \cite{cer13}.
The coverage of the uv plane achieved with the setup 6 provides low contributions of the sidelobes ($\leq$10\%\ of the primary beam) to the dirty beam.
For the rest of the setups the uv coverage is worse and large contributions of the sidelobes (up to 20--30\%\ of the primary beam) appear in the dirty beam.

The continuum comes from a point-like source located at $\alpha$=9$^{\mathrm{h}}$47$^{\mathrm{m}}$57\secp446 and $\delta$=13$^{\circ}$16$^{'}$43\arcsecp86\,(J2000), which is in good agreement with the position of IRC+10216 measured with the Very Large Array (VLA) with 40 mas resolution \citep{men12}.
We measured an intensity peak of 650 mJy\,beam$^{-1}$ with an uncertainty of $\sim$8\,\%.

The calibration of the data was performed using
CASA\footnote{\small{CASA (Common Astronomy Software Applications) is a comprehensive software package used to calibrate, image, and analyze radioastronomical data from interferometers. See:  \tt{http://casa.nrao.edu}}} 
and data analysis with
GILDAS\footnote{\small{GILDAS is a worldwide software mainly used to process, reduce, and analyze astronomical single-dish and interferometric observations. It is maintained by the Institut de Radioastronomie Millim\'{e}trique (IRAM). See: \tt{http://www.iram.fr/IRAMFR/GILDAS}}}.

\begin{figure*}[hbtp!] 
\centering
\includegraphics[width=0.98\hsize]{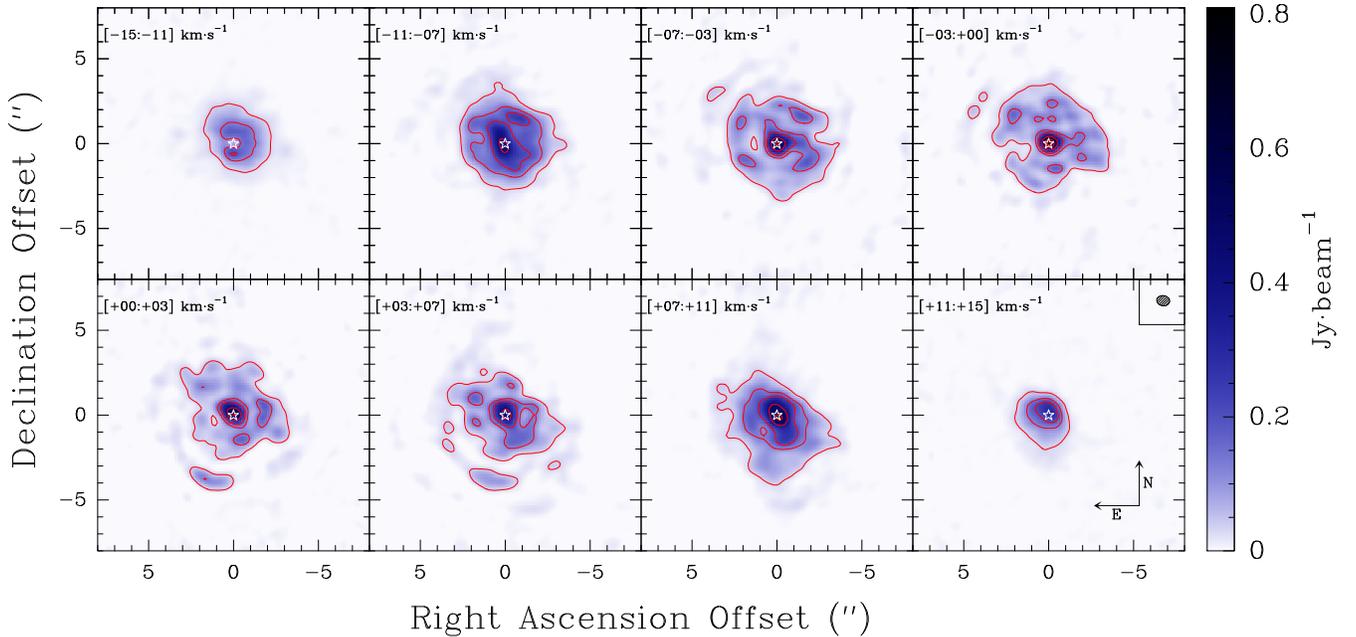}
\caption{$^{29}$SiO $J$=6 -- 5 v=0 velocity interval averaged maps. 
The velocity averaged interval is shown at the top left corner of each box.
These velocities are offset velocities with respect to the \vlsr\ of the source (-26.5\,\kms).
The position of the star is shown as a white star. 
The first contour corresponds to 5\,$\sigma$ (see Table\,\ref{tab:obs}) and the rest correspond to 15, 25, 50, 75 and 90\%\ of the maximum intensity.
The intensity scale in \jyperb\ is shown at the right edge of the figure.
The synthetic beam is represented in the top right corner of the bottom right box.
The orientation is explicitly shown in the bottom right box.}
\label{fig:29sio}
\end{figure*}
\par\bigskip

\begin{table}
\caption{Observational parameters.\label{tab:obs}}
\begin{tabular}{ccccc}
Setup \# & Frequency range & RMS & Synthetic beam & FOV$^{\dag}$ \\
         & (GHz)           & (mJy\,beam$^{-1}$) & (\arcsec$\times$\arcsec) & (\arcsec)\\
\hline
3 & 269.9--274.8 & 6  & 0.61$\times$0.47 & 23.2--22.9 \\
4 & 265.0--269.9 & 10 & 0.86$\times$0.47 & 23.7--23.2 \\
5 & 260.2--265.0 & 17 & 0.96$\times$0.47 & 24.2--23.7 \\
6 & 255.3--260.2 & 6  & 0.77$\times$0.60 & 24.6--24.2 \\
\end{tabular}
\tablecomments{$^{\dag}$Field of view calculated as $\theta$(rad)=1.22$\lambda$(mm)/D(mm), where D is the diameter of a single antenna (12\,m) and $\lambda$ is the observed wavelength.}
\end{table}

\section{Results}\label{sec:results}
\subsection{Spatial distribution of Si-bearing molecules}\label{sec:spatial}
In Figs.\,\ref{fig:si-bear} and \ref{fig:29sio}, we show maps of the emission of the lines SiC$_2$ $J_{K_{\mathrm{a}},K_{\mathrm{c}}}$=11$_{6,6}$ -- 10$_{6,5}$, $^{29}$SiO $J$=6 -- 5 and $^{29}$SiS $J$=15 -- 14 in their ground vibrational state at different offset velocities with respect to the systemic velocity of the source.

% SiC2
SiC$_2$ (Fig.\,\ref{fig:si-bear}) displays a central component elongated in the NE--SW direction with a size of $\sim$4--5\arcsec\ along the nebular axis and $\sim$3--4\arcsec\ in the perpendicular direction.
The elongation is also observed in the $^{29}$SiO emission (see below) and in the SiO and SiC$_2$ maps by \cite{fon14} where the authors invoke a possible bipolar outflow to explain it.
At the systemic velocity of the source and at +6\,\kms\ offset, a ringlike, clumpy and weak component is seen at $\sim$10--11\arcsec\ from the central star. 
The angular distance between the position of the star and the ring structure, considering a distance of $\sim$130\,pc to the star from us \citep{gro12}, corresponds to a linear distance of $\sim$2$\times$10$^{16}$\,cm.
This ringlike component is consistent with the peak abundance of SiC$_2$ in the outer envelope of IRC+10216 reported in \cite{luc95} and the chemical model of \cite{cer10}.
Although, this ringlike structure is probably filtered in our data given the shortest baselines used.  
Between the central and the ringlike structure, emission of SiC$_{2}$ is either very low or absent \citep{luc95}.
Finally, the redshifted emission (+12\,\kms) displays a quasi-circular distribution with a diameter of $\sim$5--6\arcsec. 
These brightness distributions could be interpreted as the SiC$_{2}$ is formed in regions close to the star, then it condenses onto the dust grains, and eventually it reappears at the outer shells of the CSE, perhaps as a hollow shell, as the consequence of the interaction between the UV Galactic radiation field and the CSE \citep{luc95,fon14}.

% SiO
For $^{29}$SiO (Figs.\,\ref{fig:si-bear} and \ref{fig:29sio}), the bulk of the emission arises from a compact central component with a size of 2\arcsec.
This line also displays an extended and clumpy distribution, elongated in the NE--SW and with a size of $\sim$6--7\arcsec.
At velocities close to the terminal expansion velocity, $\sim$14.5\,\kms \citep{cer00}, the brightness distribution is elongated in the NE--SW direction with a size of $\sim$3--4\arcsec.
The blueshifted emission at -13\,\kms\ displays a decrease just in front of the star, which can be interpreted as self-absorption and probably absorption of the continuum emission mostly coming from the star.
There is no conclusive explanation for the elongation; nevertheless, some authors pointed out that it could evidence the presence of a bipolar outflow in the CSE \citep{fon14}.
The possible presence of a binary companion to CWLeo could also play a decisive role in this scenario \citep{gue93,cer15}.

The observed brightness distributions of the vibrationally excited SiS lines are expected to be compact and centered on the star since the involved levels are excited at the high temperatures prevailing close to the star.
The SiS $J$=15 -- 14 v=0 line, which displays maser emission \citep{fon06}, shows a circular brightness distribution with a diameter of $\sim$2\arcsec\ at the systemic velocity of the source. 
For the SiS $J$=15 -- 14 v$\geq$1 lines, the observed distributions are not spatially resolved.
Fig.\,\ref{fig:si-bear} shows the emission of the $^{29}$SiS $J$=15 -- 14 v=0 line, which displays a brightness distribution of a compact source surrounding the central star with a diameter of $\sim$2\arcsec. 
A large-scale artificial modulation can be seen below the 25$\sigma$ level.
Since it is related to the visibilities at short baselines, we do not expect it to modify the shape and flux of the compact central component of the brightness distribution.
Also, the quality of the several SiS and $^{29}$SiS maps is affected owing to the low uv coverage and non-neglible contribution of the sidelobes for the setups 3--5 (see Table\ref{tab:obs}).
The lines that lie in the range covered by the setup 6, are those of high vibrationally excited states (e.g. SiS $J$=15 -- 14 v$\geq$10, $^{29}$SiS $J$=15 -- 14 v$\geq$6) which are tentative and spatially unresolved.

\begin{figure*}[hbtp!] 
\centering
\includegraphics[width=0.80\hsize]{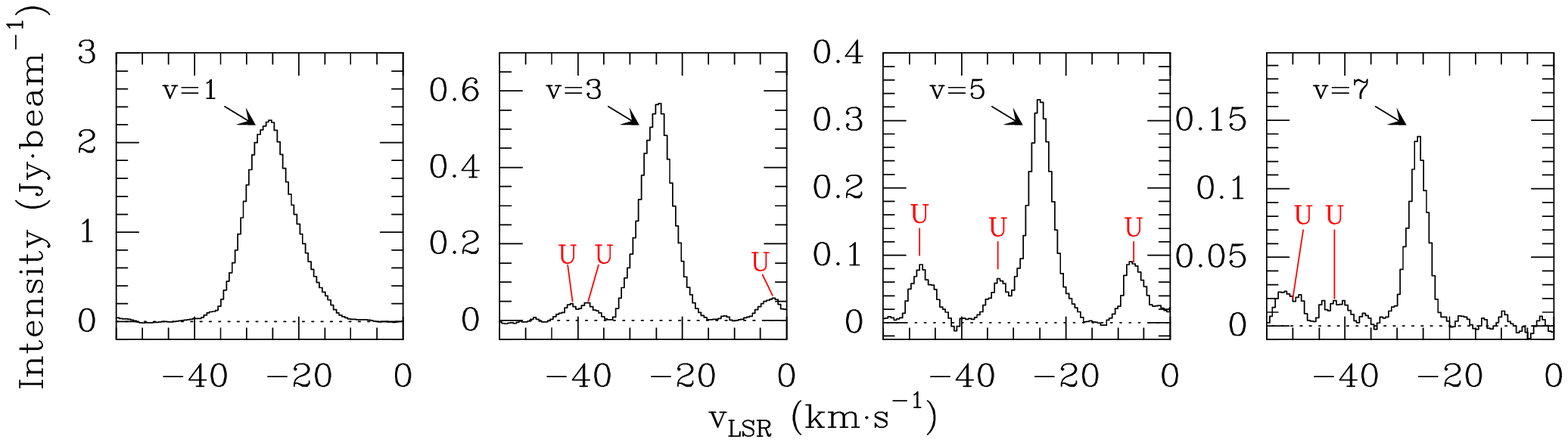}\label{a}
\par\medskip
\includegraphics[width=0.80\hsize]{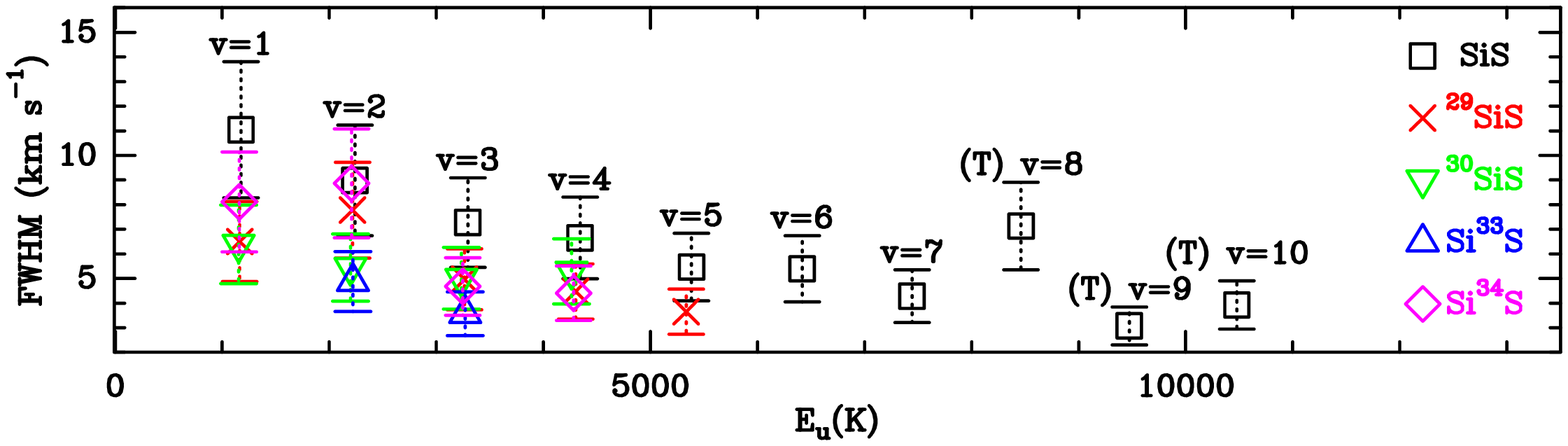}\label{b}
\par\medskip
\includegraphics[width=0.80\hsize]{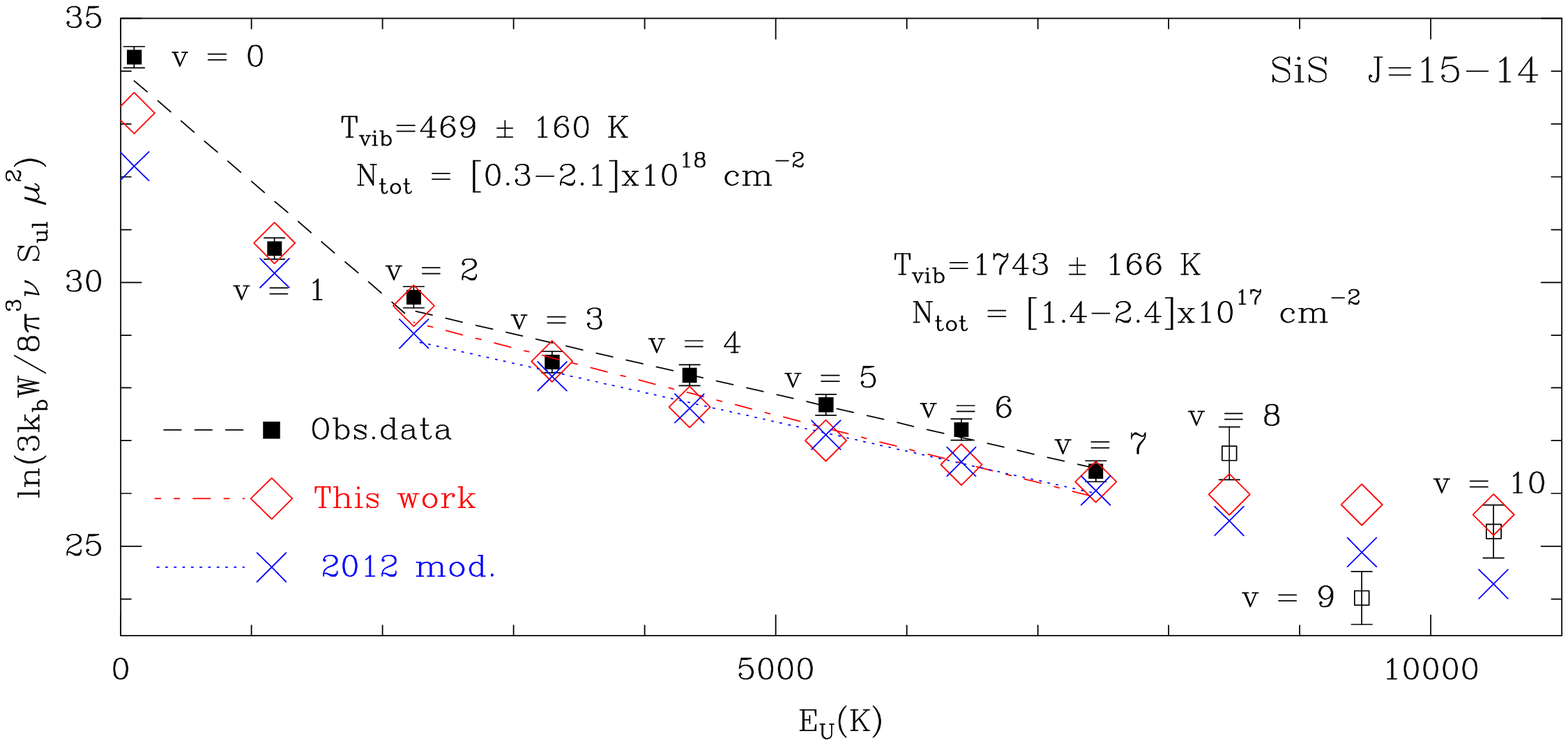}\label{c}
% \vspace{0.2pt}
\caption{
{\it{(Top)}} Central pixel spectra of the SiS $J$=15 -- 14 v=1, 3, 5 and 7 lines. Unidentified lines are marked with an U.
Baselines are indicated in each box with a dashed line.
{\it{(Middle)}} FWHM of the v$\geq$1 detected lines of SiS isotopologues in the central pixel versus the upper energy of the transition.
The tentative lines are marked with a (T).
We adopted a 25\%\ value of the total measure for the uncertainties, which takes into account uncertainties of the calibration, statistical errors, and the contribution of the sidelobes of the dirty beam.
{\it{(Bottom)}} Vibrational diagram of the SiS $J$=15 -- 14 lines.
The vibrational quantum number of each line is shown.
The errorbars represent a 20\% of the integrated intensity of the line and a 50\% for the v=8, 9 and 10 lines.
The values derived for the vibrational temperature and column density are shown with their formal errors for two fits: v=0--2 and v=3--7.
The LVG models (see \S\,\ref{sec:sisvib}) are plotted with a rhombus and a cross.
}
\label{fig:sis_all}
\end{figure*}

\subsection{Vibrationally excited SiS}\label{sec:sisvib}
The Cycle\,0 observations with ALMA allowed us to detect several $J$=15 -- 14 lines of high vibrationally excited states of SiS isotopologues, in particular, up to v=7 for the main isotopologue (see Fig.\,\ref{fig:sis_all}).
The SiS $J$=15 -- 14 v=8 and v=10 lines are probably blended with unidentified lines, so we consider them to be tentative.
Additionally, the SiS $J$=15 -- 14 v=9 line is considered tentatively detected because even though its FWHM measured seems to follow the trend shown in Fig.\,\ref{fig:sis_all}, its measured integrated intensity is underestimated considering the population diagram of Fig.\,\ref{fig:sis_all}.
All these lines display a compact unresolved emission peaking at the central star.
The FWHM of the lines measured from the spectra at the stellar position decreases with increasing upper level energy (see Fig.\,\ref{fig:sis_all}).
We verified this behavior for SiS, $^{29}$SiS, $^{30}$SiS, Si$^{33}$S and Si$^{34}$S. 
In the dust formation region, the gas displays a velocity gradient as a function of the radial distance to the star, i.e., the closer to the star the lower the expansion velocity \citep[][and references therein]{agu12}. 
Hence, those lines involving higher vibrational states, which are excited in inner and warmer regions, are narrower. 
Thermal broadening for the SiS lines excited in the dust formation region is $\sim$1\,\kms, and thus this mechanism could only account partially for the FWHM variation of the lines.

% \par\bigskip
We analyzed the excitation conditions of SiS with the rotational diagram technique \citep{gol99} using the spectra at the stellar position (see Fig.\,\ref{fig:sis_all}).
We considered two different linear trends for the observational data: one for the transitions with E$_{up}$$\lesssim$2500\,K (i.e. v=0, 1 and 2), and a different trend for the rest (i.e. v=3--7).
The v$\geq$8 lines are excluded from the fit.
The values derived from the v=0 to v=2 fit are uncertain owing to the maser nature of the SiS $J$=15 -- 14 v=0 line and also due to possible optically thick emission.
Both data series show a linear behavior consistent with a single vibrational (excitation) temperature for each of the fits.
The vibrational temperature derived from the fit involving the transitions in high-energy vibrational states is higher than the temperature derived for the low vibrational levels.
Therefore, the emission produced by SiS transitions in high-energy vibrational states can only arise from regions close to the photosphere.
This result is similar to the one obtained by \cite{cer11} for HCN.

We used a large velocity gradient (LVG) code to model the SiS emission \citep{cer12}.
Further details about the spectroscopic data used in the calculations are given in \S\,\ref{sec:vibmom}.
The SiS collisional data were taken from \cite{tob08} and extrapolated to high rovibrational levels.
We adopted a distance to the star of $\sim$130\,pc, an effective temperature of $\sim$2330\,K and a stellar radius of $\sim$4$\times$10$^{13}$\,cm as input for our model \citep{cer00,mon00,gro12}.
We used two different models:
($i$) the model of \cite{agu12}, which was used to reproduce the molecular abundances in the inner layers of IRC+10216, that we call the ``2012 model'', and,
($ii$) this work, which is a modification of the model ($i$).
We modified the H$_2$ density, decreasing it by a factor $\sim$2, as described in \cite{cer13} (this was used to reproduce the dust nucleation zone, 1--10\,\rs, of IRC+10216).
Additionally, the SiS abundance in the dust nucleation zone needs to be balanced with a similar increase (factor of $\sim$2) to avoid an underestimation of the SiS emission.
From these models (see Fig.\,\ref{fig:sis_all}) we obtain a good agreement with the vibrational temperature derived from the vibrational diagram and moderate discrepancies with the total column density within a factor of $\sim$2.
These discrepancies in the column density may be explained by the dilution due to the size of the emitting region compared to the half power beam width of the synthetic beam, which would increase the optical depth of the lines.
The size of the emitting region should decrease with the vibrational state owing to the energies needed to excite those lines.
With our models, we found moderate to high optical depths $\tau$(v=1)$\sim$10 to $\tau$(v=4)$\sim$0.8 for the abundance profile ($i$) and $\tau$(v=1)$\sim$47 to $\tau$(v=8)$\sim$1.0 for the abundance profile ($ii$).

\subsubsection{SiS potential energy function and vibrational dipole moment}\label{sec:vibmom}
The potential energy function used to describe the internuclear motions of the SiS isotopologues is a Born--Oppenheimer (BO) potential properly extended 
to accommodate Born--Oppenheimer breakdown (BOB) corrections. The effective potential has the form \citep{Campbell93,Ram97,Dulick98,Coxon00}

\begin{equation}
V^{eff}(r) = V^{\rm{BO}}(r) + \frac{V_{A}(r)}{M_{A}} + \frac{V_{B}(r)}{M_{B}} -\frac{\hbar^{2}}{2\mu}\left[1 + q(r)\right]\frac{J(J+1)}{r^{2}} 
\label{pefdmm1}
\end{equation}
where $M_{A}$ and $M_{B}$ are the silicon and sulfur atomic masses and $\mu$ is the reduced mass. The BO potential is given by 
\begin{equation}
V^{\rm{BO}}(r) = D_{e}\left[\frac{1-e^{-\beta(r)}}{1-e^{-\beta(\infty})}\right]^{2}
\label{pefdmm2}
\end{equation}
where
\begin{equation}
\beta(r) = z\sum_{i=0}^{4} \beta_{i} z^{i}
\label{pefdmm3}
\end{equation}
\begin{equation}
\beta(\infty) = \sum_{i=0}^{4} \beta_{i} 
\label{pefdmm4}
\end{equation}
and
\begin{equation}
z=\frac{r-r_{e}}{r+r_{e}}
\label{pefdmm5}
\end{equation}
and the BOB potential and centrifugal correction terms are represented by the power expansions %\citep{Campbell93,Dulick98,Coxon00}
\begin{equation}
V_{A}(r) = \sum_{i=1}u_{i}^{A}z^{i}
\label{pefdmm6}
\end{equation}
\begin{equation}
V_{B}(r) = \sum_{i=1}u_{i}^{B}z^{i}
\label{pefdmm7}
\end{equation}
\begin{equation}
q(r) = M_{A}^{-1}\sum_{i=1}q_{i}^{A}z^{i} + M_{B}^{-1}\sum_{i=1}q_{i}^{B}z^{i} 
\label{pefdmm8}
\end{equation}

The potential parameters were obtained by nonlinear least squares fitting to the observed infrared and microwave line positions of the SiS isotopologues up to v=12. 
The final data set included a total of 2863 lines, 414 rotational transitions \citep{Tiemann1972,Sanz2003,Muller2007} and 2449 rovibrational transitions \citep{Birk1972,Frum1990}. 
Mass-independent Dunham coefficients, U$_{ij}$, have been derived by \cite{Muller2007}.
The fit to the potential energy function was performed using the Levenberg--Marquardt algorithm \citep{Levenberg44,Marquardt63} to minimize the $\chi^{2}$ function, with the line positions weighted by the square of the experimental uncertainties. 
The rovibrational energy levels of the isotopologues needed to calculate the line positions were computed by solving the radial Schr\"odinger equation using the variational method of \cite{Harris65} along with harmonic-type basis functions. 
The potential parameters obtained in the fit are given in the comments of the Table\,\ref{vdip}.
The final $\chi^{2}$ value was 1.767. 
The unweighted standard deviations for the rotational and, vibrational-rotational line positions were 0.0229\,MHz and 0.000657\,cm$^{-1}$, respectively.

The dipole moment function used for SiS was determined semiempirically by \cite{LopezPineiro87}, and the dipole moment matrix elements were computed up to v=4. 
In Table\,\ref{vdip} we provide them up to v=12. 
The agreement between our calculations and those of \cite{LopezPineiro87} for v$\le$4 is excellent.

\subsection{Other Si-bearing species}\label{sec:others}
SiC was detected in IRC+10216 by \cite{cer89} with line profiles indicating that the molecule was produced in an external shell, probably as a product of the photodissociation of SiC$_2$.
Our observations did not cover any frequency range where SiC lines could arise; however, lines of $^{29}$SiC and SiC v=1 lie in this range, but they were not detected.
For a temperature in the photosphere of 2300\,K a significant number of SiC molecules could be in the v=1 state (E$_{\mathrm{up}}$$\sim$1400\,K) and higher vibrational levels.
We derived an upper limit to the SiC column density of 4.4$\times$10$^{14}$\,cm$^{-2}$ from the $^{29}$SiC upper limit, where we used an isotopic $^{28}$Si/$^{29}$Si ratio of 20 \citep{cer89,cer91}.
Hence, SiC$_2$ is the main carrier of SiC bonds in the gas phase in the dust formation zone of IRC+10216.

\section{Conclusion}\label{sec:conc}
ALMA has proved to be an outstanding tool to study the molecular emission from CSEs of evolved stars, even at the early stages of its development.
In particular, ALMA allowed us to detect SiS rotational lines in high-energy vibrational states that have been analyzed to constrain the physical conditions of the innermost shells of IRC+10216.
We found that these lines should be excited in regions close to the photosphere of IRC+10216.
It also has served to unveil the different brightness distributions of Si-bearing molecules.
We expect that future ALMA science, with its full suite of capabilities ready for the next observation cycle, would give us the chance to map the brightness distributions of these Si-bearing molecules in greater detail, allowing us to understand their formation mechanisms.
% \newpage

%% ACKNOWLEDGMENTS: ----------------------------------------------------------
\begin{acknowledgments}
% {\bf{ACKNOWLEDGMENTS}}
We thank the Spanish MINECO/MICINN for funding support through grants AYA2009-07304, AYA2012-32032, 
the ASTROMOL Consolider project CSD2009-00038 and the European Research Council (ERC Grant 610256: NANOCOSMOS).
\end{acknowledgments}
%% FACILITIES: ----------------------------------------------------------

%% in parentheses, after the keyword, but they will not be verified.

% {\it Facilities:} \facility{ALMA}. 
% BIBLIOGRAPHY -----------------------------------------------------------

% \end{document}

\clearpage
\begin{turnpage}
% \begin{landscape}
% \topskip-20pt
\begin{table}
\begin{center}
% \setlength\LTcapwidth{\textwidth} % default: 4in (rather less than \textwidth...)
% \setlength\LTleft{0pt}            % default: \parindent
% \setlength\LTright{0pt}           % default: \fill
% \begin{minipage}{28cm}
\caption{Calculated vibrational dipole moment matrix elements (D) for $^{28}$Si$^{32}$S.}
% \vspace{-0.5cm}
\label{vdip}
\tabcolsep 0.5mm
%\tabcolsep 1mm
%\tabcolsep 5mm
% \begin{small}
\resizebox{22cm}{!}{
\begin{tabular}{cccccccccccccc} 
 $v'$/$v"$ &   0 & 1  & 2 & 3  & 4 & 5 & 6 & 7 & 8 & 9 & 10 & 11 & 12 \\
\hline
 0 & 1.7420    &           &          &          &          &          &          &          &          & & & &  \\
 1 &-0.1348    & 1.7658    &          &          &          &          &          &          &          & & & &  \\   
 2 &-0.6333$\times$10$^{-2}$ &-0.1907    & 1.7896   &          &          &          &          &          &          & & & &  \\  
 3 & 0.4068$\times$10$^{-3}$ & 0.1101$\times$10$^{-1}$ & 0.2336   & 1.8135   &          &          &          &          &          & & & &  \\    
 4 &-0.3379$\times$10$^{-4}$ &-0.8184$\times$10$^{-3}$ &-0.1564$\times$10$^{-1}$& 0.2697   & 1.8375   &          &          &          &          & & & &  \\  
 5 &-0.3359$\times$10$^{-5}$ &-0.7613$\times$10$^{-4}$ &-0.1302$\times$10$^{-2}$& 0.2027$\times$10$^{-1}$&-0.3016   & 1.8615   &          &          &          & & & &  \\  
 6 & 0.3804$\times$10$^{-6}$ &-0.8306$\times$10$^{-5}$ & 0.1329$\times$10$^{-3}$&-0.1852$\times$10$^{-2}$&-0.2493$\times$10$^{-1}$& 0.3304   & 1.8857   &          &          & & & &  \\  
 7 & 0.4752$\times$10$^{-7}$ & 0.1018$\times$10$^{-5}$ & 0.1569$\times$10$^{-4}$&-0.2046$\times$10$^{-3}$& 0.2465$\times$10$^{-2}$& 0.2961$\times$10$^{-1}$&-0.3568   & 1.9100   &          & & & &  \\               
 8 &-0.6394$\times$10$^{-8}$ &-0.1363$\times$10$^{-6}$ &-0.2061$\times$10$^{-5}$& 0.2587$\times$10$^{-4}$&-0.2916$\times$10$^{-3}$&-0.3137$\times$10$^{-2}$& 0.3433$\times$10$^{-1}$& 0.3814   & 1.9343   & & & &  \\
 9 &-0.9087$\times$10$^{-9}$ &-0.1952$\times$10$^{-7}$ &-0.2934$\times$10$^{-6}$& 0.3612$\times$10$^{-5}$&-0.3918$\times$10$^{-4}$&-0.3943$\times$10$^{-3}$& 0.3866$\times$10$^{-2}$& 0.3909$\times$10$^{-1}$&-0.4045   & 1.9587& & &  \\
10 & 0.1339$\times$10$^{-9}$ & 0.2936$\times$10$^{-8}$ & 0.4441$\times$10$^{-7}$&-0.5435$\times$10$^{-6}$& 0.5779$\times$10$^{-5}$& 0.5594$\times$10$^{-4}$&-0.5130$\times$10$^{-3}$&-0.4649$\times$10$^{-2}$& 0.4388$\times$10$^{-1}$& 0.4262& 1.9833& &  \\ 
11 & 0.2005$\times$10$^{-10}$& 0.4562$\times$10$^{-9}$ & 0.7033$\times$10$^{-8}$&-0.8654$\times$10$^{-7}$& 0.9143$\times$10$^{-6}$& 0.8674$\times$10$^{-5}$&-0.7649$\times$10$^{-4}$&-0.6482$\times$10$^{-3}$& 0.5485$\times$10$^{-2}$& 0.4871$\times$10$^{-1}$&-0.4469& 2.0079&  \\
12 &-0.2954$\times$10$^{-11}$&-0.7182$\times$10$^{-10}$&-0.1147$\times$10$^{-8}$& 0.1437$\times$10$^{-7}$&-0.1525$\times$10$^{-6}$&-0.1437$\times$10$^{-5}$& 0.1241$\times$10$^{-4}$& 0.1011$\times$10$^{-3}$&-0.8003$\times$10$^{-3}$&-0.6373$\times$10$^{-2}$& 0.5358$\times$10$^{-1}$& 0.4667& 2.0326 \\ 
\hline\hline                                                                    
\end{tabular} 
}
\tablecomments{
Internuclear potential energy parameters for the SiS molecule:
D$_{e}$=51993.7274\,cm$^{-1}$, 
r$_{e}$=1.929260274\,\AA,
$\beta_{0}$=5.9671641,
$\beta_{1}$=5.8793706,
$\beta_{2}$=9.3200373,
$\beta_{3}$=18.402479,
$\beta_{4}$=18.451287,
u$_{1}^{\rm{Si}}$=-749.1309\,cm$^{-1}$\,amu,
u$_{2}^{\rm{Si}}$=6844.274\,cm$^{-1}$\,amu,
u$_{1}^{\rm{S}}$=-1003.321\,cm$^{-1}$\,amu,
u$_{2}^{\rm{S}}$=7128.960\,cm$^{-1}$\,amu.
}
% \end{small}
% \end{minipage}
\end{center}
\end{table}      
% \end{landscape}
% \end{changemargin} 
% \restoregeometry
\end{turnpage}
% \clearpage
% \global\pdfpageattr\expandafter{\the\pdfpageattr/Rotate 90}

\begin{thebibliography}{}
\bibitem[Ag{\'u}ndez et 
al.(2012)]{agu12} Ag{\'u}ndez, M., Fonfr{\'{\i}}a, J.~P., Cernicharo, J., et al.\ 2012, \aap, 543, AA48 

\bibitem[Bieging 
\& Tafalla(1993)]{bie93} Bieging, J.~H., \& Tafalla, M.\ 1993, \aj, 105, 576 

\bibitem[Birk et al. (1972)]{Birk1972}Birk H., and Jones H., 1972, Chem. Phys. Lett. 175, 536.

\bibitem[Campbell et al. (1993)]{Campbell93} Campbell, J.~M., Dulick, M., Klapstein, D., et al., 1993, J. Chem. Phys., 99, 8379

\bibitem[Castro-Carrizo et 
al.(2001)]{cas01} Castro-Carrizo, A., Lucas, R., Bujarrabal, V., Colomer, F., \& Alcolea, J.\ 2001, \aap, 368, L34 

\bibitem[Cernicharo et al.(1989)]{cer89} Cernicharo, J., 
Gottlieb, C.~A., Guelin, M., Thaddeus, P., 
\& Vrtilek, J.~M.\ 1989, \apjl, 341, L25 

\bibitem[Cernicharo et 
al.(1991)]{cer91} Cernicharo, J., Guelin, M., Kahane, C., Bogey, M., \& Demuynck, C.\ 1991, \aap, 246, 213 

\bibitem[Cernicharo et 
al.(2000)]{cer00} Cernicharo, J., Gu{\'e}lin, M., \& Kahane, C.\ 2000, \aaps, 142, 181 

\bibitem[Cernicharo et 
al.(2010)]{cer10} Cernicharo, J., Waters, L.~B.~F.~M., Decin, L., et al.\ 2010, \aap, 521, LL8 

\bibitem[Cernicharo et 
al.(2011)]{cer11} Cernicharo, J., Ag{\'u}ndez, M., Kahane, C., et al.\ 2011, \aap, 529, LL3 

\bibitem[Cernicharo(2012)]{cer12} Cernicharo, J.\ 2012, EAS 
Publications Series, 58, 251 

\bibitem[Cernicharo et al.(2013)]{cer13} Cernicharo, J., 
Daniel, F., Castro-Carrizo, A., et al.\ 2013, \apjl, 778, LL25 

\bibitem[Cernicharo et 
al.(2015)]{cer15} Cernicharo, J., Marcelino, N., Ag{\'u}ndez, M., \& Gu{\'e}lin, M.\ 2015, \aap, 575, AA91 

\bibitem[Coxon \& Hajigeorgiou (2000)]{Coxon00}  Coxon, J.~A.,  and Hajigeorgiou, P.~G., 2000, J. Mol. Spectrosc., 203, 49

\bibitem[Dulick et al. (1998)]{Dulick98} Dulick, M.,  Zhang, K.~Q., Guo, B., and  Bernath, P.~F., 1998, J. Mol. Spectrosc., 188, 14

\bibitem[Fonfr{\'{\i}}a Exp{\'o}sito et al.(2006)]{fon06} 
Fonfr{\'{\i}}a Exp{\'o}sito, J.~P., Ag{\'u}ndez, M., Tercero, B., Pardo, 
J.~R., \& Cernicharo, J.\ 2006, \apjl, 646, L127 

\bibitem[Fonfr{\'{\i}}a et al.(2014)]{fon14} Fonfr{\'{\i}}a, 
J.~P., Fern{\'a}ndez-L{\'o}pez, M., Ag{\'u}ndez, M., et al.\ 2014, \mnras, 
445, 3289 

\bibitem[Frum et al. (1990)]{Frum1990} Frum C.I., Engleman Jr. R., Bernath P.F., 1990, J. Chem. Phys.,   93, 5457

\bibitem[Goldsmith 
\& Langer(1999)]{gol99} Goldsmith, P.~F., \& Langer, W.~D.\ 1999, \apj, 517, 209 

\bibitem[Groenewegen et 
al.(2012)]{gro12} Groenewegen, M.~A.~T., Barlow, M.~J., Blommaert, J.~A.~D.~L., et al.\ 2012, \aap, 543, LL8 

\bibitem[Guelin et 
al.(1993)]{gue93} Guelin, M., Lucas, R., \& Cernicharo, J.\ 1993, \aap, 280, L19 

\bibitem[Harris et al. (1965)]{Harris65} Harris, D.~O., Engerholm, G.~G., and Gwinn, W.~D., 1965., J. Chem. Phys., 43, 1515

% \bibitem[Jeffers et 
% al.(2014)]{jef14} Jeffers, S.~V., Min, M., Waters, L.~B.~F.~M., et al.\ 2014, \aap, 572, AA3 

\bibitem[Levenberg (1944)]{Levenberg44} Levenberg, K., 1944, Quart. Appl. Math., 2, 164

\bibitem[Lucas et 
al.(1995)]{luc95} Lucas, R., Gu{\'e}lin, M., Kahane, C., Audinos, P., \& Cernicharo, J.\ 1995, \apss, 224, 293 

% % \bibitem[Lucas(1997)]{luc97} Lucas, R.\ 1997, \apss, 251, 247 

\bibitem[Marquardt (1963)]{Marquardt63}  Marquardt, D.~W., 1963, J. Soc. Ind. Appl. Math., 11, 431

\bibitem[Menten et 
al.(2012)]{men12} Menten, K.~M., Reid, M.~J., Kami{\'n}ski, T., \& Claussen, M.~J.\ 2012, \aap, 543, AA73

\bibitem[Monnier et al.(2000)]{mon00} Monnier, J.~D., Danchi, 
W.~C., Hale, D.~S., et al.\ 2000, \apj, 543, 861 

\bibitem[Morris et al.(1975)]{mor75} Morris, M., Gilmore, W., 
Palmer, P., Turner, B.~E., \& Zuckerman, B.\ 1975, \apjl, 199, L47 %1st detection of SiS

\bibitem[M\"uller et al. (2007)]{Muller2007} M\"uller, H.S.P., McCarthy, M.C., Bozzocchi, L., et al., 2007, Chem. Phys. Phys. Chem., 9, 1579

\bibitem[Olofsson et 
al.(1982)]{olo82} Olofsson, H., Johansson, L.~E.~B., Hjalmarson, A., \& Nguyen-Quang-Rieu 1982, \aap, 107, 128 

\bibitem[Pi\~neiro et al. (1987)]{LopezPineiro87} Pi\~neiro, A.~L.~L., Tipping, R.~H., and Chackerian, J.~C., 1987, J. Mol. Spectrosc., 125, 184

\bibitem[Ram et al. (1997)]{Ram97} Ram, R.~S., Dulick, M., Guo, B., et al., 1997, J. Mol. Spectrosc., 183, 360

\bibitem[Sanz et al. (2003)]{Sanz2003}Sanz M.E., McCarthy M.C., Thaddeus P., 2003, J. Chem. Phys., 119, 11715

\bibitem[Sch{\"o}ier et al.(2006)]{sch06} Sch{\"o}ier, F.~L., 
Fong, D., Olofsson, H., Zhang, Q., \& Patel, N.\ 2006, \apj, 649, 965 

\bibitem[Tieman et al. (1972)]{Tiemann1972} Tiemann E., Renwanz E., Hoeft J., Torring T., 1972, Z. Naturforsch. 27a, 1566

\bibitem[Tobo{\l}a et al.(2008)]{tob08} Tobo{\l}a, R., Lique, 
F., K{\l}os, J., 
\& Cha{\l}asi{\'n}ski, G.\ 2008, Journal of Physics B Atomic Molecular Physics, 41, 155702 

\bibitem[Tsuji(1973)]{tsu73} Tsuji, T.\ 1973, \aap, 23, 411 
\end{thebibliography}
\end{document}